\documentclass[letterpaper, 10 pt, conference]{ieeeconf}  % Comment this line out if you need a4paper

\IEEEoverridecommandlockouts                              % This command is only needed if 
\usepackage{color,graphicx}
\usepackage{amsmath,amsfonts,bm}

\def\eqref#1{equation~\ref{#1}}
\def\1{\bm{1}}

\def\vf{{\bm{f}}}

\def\vh{{\bm{h}}}

\def\vn{{\bm{n}}}

\def\vu{{\bm{u}}}
\def\vv{{\bm{v}}}
\def\vw{{\bm{w}}}
\def\vx{{\bm{x}}}
\def\vy{{\bm{y}}}
\def\vz{{\bm{z}}}

\def\mA{{\bm{A}}}

\def\mU{{\bm{U}}}
\def\mV{{\bm{V}}}

\def\mX{{\bm{X}}}
\def\mY{{\bm{Y}}}
\def\mZ{{\bm{Z}}}

\DeclareMathAlphabet{\mathsfit}{\encodingdefault}{\sfdefault}{m}{sl}
\SetMathAlphabet{\mathsfit}{bold}{\encodingdefault}{\sfdefault}{bx}{n}

\title{\LARGE \bf
Dual Parametric and State Estimation for Partial Differential Equations
}

\author{Saviz Mowlavi$^{1}$ and Mouhacine Benosman$^{2}$% <-this % stops a space
\thanks{$^{1,2}$Mitsubishi Electric Research Laboratories,
        Cambridge, MA02139, USA.
        {\tt\small mowlavi@merl.com, benosman@merl.com}}%
}

\begin{document}

\maketitle
\thispagestyle{empty}
\pagestyle{empty}

%%%%%%%%%%%%%%%%%%%%%%%%%%%%%%%%%%%%%%%%%%%%%%%%%%%%%%%%%%%%%%%%%%%%%%%%%%%%%%%%
\begin{abstract}

Designing estimation algorithms for systems governed by partial differential equations (PDEs) such as fluid flows is challenging due to the high-dimensional and oftentimes nonlinear nature of the dynamics, as well as their dependence on unobserved physical parameters. In this paper, we propose two different lightweight and effective methodologies for real-time state estimation of PDEs in the presence of parametric uncertainties. Both approaches combine a Kalman filter with a data-driven polytopic linear reduced-order model obtained by dynamic mode decomposition (DMD). Using examples involving the nonlinear Burgers and Navier-Stokes equations, we demonstrate accurate estimation of both the state and the unknown physical parameter along system trajectories corresponding to various physical parameter values.

\end{abstract}

%%%%%%%%%%%%%%%%%%%%%%%%%%%%%%%%%%%%%%%%%%%%%%%%%%%%%%%%%%%%%%%%%%%%%%%%%%%%%%%%
\section{INTRODUCTION}

Active control of fluid flows has many potential benefits, from drag reduction in aircrafts and ships to improved efficiency of heating and air-conditioning systems, among many other examples \cite{brunton2015closed}. But real-time feedback control requires inferring the state of the system from sparse measurements using a state estimation algorithm, which typically relies on a model for the underlying dynamics \cite{simon2006optimal}. Among estimation algorithms, the Kalman filter is by far the most well-known thanks to its optimality for linear systems, which has led to its widespread use in numerous applications \cite{kalman1960,zarchan2005progress}. However, systems such as fluid flows are governed by partial differential equations (PDEs) which, when discretized, yield high-dimensional and oftentimes nonlinear dynamical models with hundreds or thousands of state variables. These high-dimensional models are too expensive to integrate with common state estimation techniques, especially in the context of embedded systems. Thus, for control purposes, a common practice is to design state estimators from a reduced-order model (ROM) of the system, in which the underlying dynamics are projected to a low-dimensional subspace that is computationally tractable \cite{barbagallo2009closed,rowley2017}. In particular, recent papers have demonstrated the efficacy of combining a data-driven ROM constructed using dynamic mode decomposition (DMD) with a Kalman filter to estimate unsteady fluid flows using sparse sensor measurements in the absence of model uncertainties \cite{gomez2019data,tsolovikos2020estimation}. 

%We propose to add to these interesting works by focusing on the dual parameter-state estimation problem. In this context, we propose to learn the best estimate of the uncertain parameter and update the state estimator in parallel to achieve optimal state and parametric estimation for fluid flows.

The dynamics of physical systems such as fluid flows is oftentimes sensitive to various physical parameters such as wind speed, viscosity, etc. When these parameters are unknown, the accuracy of the estimate is adversely impacted by the uncertainty in the model and, thus, the ROM. To deal with these model uncertainties, an extension of the previously referenced studies was recently proposed, wherein a bank of local DMD models for multiple parameter values was utilized in a multiple model Kalman filter framework, in which the estimate is a weighted average of individual estimates produced by independent Kalman filters running separately for each parameter value \cite{tsolovikos2022multiple}. This latter framework belongs to a broader class of multiple model techniques addressing parameter uncertainties \cite{orjuela2008state,adeniran2016modeling}, which also include polytopic models where a single state or estimate is produced by a weighted average of the local models themselves \cite{takagi1985fuzzy,apkarian1995self,angelis2003system,fujimori2006model}. 

In this paper, we adopt the polytopic modeling approach to formulate an adaptive estimator for systems governed by PDEs in the presence of parameter uncertainties. Specifically, we construct a data-driven polytopic ROM from a bank of local DMD models for multiple parameter values, which we combine with a joint or modular Kalman filter algorithm to simultaneously learn online an estimate for both the state and the unknown parameter. We compare the proposed dual parameter-state estimation methods with a classical robust estimation approach, where a single global DMD model is computed from a wide range of the uncertain parameter, without online adaptation of the parameter estimate. The performance of these various estimation algorithms are tested on two PDE systems, the Burgers equation and the Navier-Stokes equations.

The remaining of the paper is organized as follows: Section \ref{sec:ProblemFormulation} is dedicated to the statement of the estimation problem. In Section \ref{sec:ROM}, we recall the DMD method and present two ROM approaches, a robust ROM and polytopic multi-ROMs approach. In Section \ref{sec:estimation}, the proposed dual parameter-state estimators are introduced. The numerical results are reported in Section \ref{sec:Simulations}. Finally, conclusions and future works are discussed in Section \ref{sec:Conclusions}.

\section{PROBLEM FORMULATION}
\label{sec:ProblemFormulation}

Consider the parametric discrete-time nonlinear system
\begin{subequations}
\begin{align}
\vz_{k+1} &= \vf(\vz_k; p), \label{eq:FODynamics1} \\
\vy_k &= \vh(\vz_k,\vn_k),
\end{align} \label{eq:FODynamics}%
\end{subequations}
where $\vz_k \in \mathbb{R}^n$ and $\vy_k \in \mathbb{R}^m$ are respectively the state and measurement at time $k$,  $p \in [p_\mathrm{min},p_\mathrm{max}] \subset \mathbb{R}$ is an uncertain physical parameter, $\vf: \mathbb{R}^n \times \mathbb{R} \rightarrow \mathbb{R}^n$ is a nonlinear map from current state and parameter to next state, $\vn_k \in \mathbb{R}^m$ is observation noise, and $\vh: \mathbb{R}^n \times \mathbb{R}^m \rightarrow \mathbb{R}^m$ is a nonlinear map from current state and noise to measurement. In this study, we assume that the dynamics given in (\ref{eq:FODynamics}) are obtained from \textcolor{black}{a high-fidelity} numerical discretization of a nonlinear PDE, which typically requires a large number $n$ of continuous state variables (on the order of at least a few hundreds). Nonetheless, our work is applicable to any high-dimensional nonlinear system of the form (\ref{eq:FODynamics}), so long as a linear model can provide a useful approximation of the system at hand.

Here, we will focus on the post-transient dynamics of (\ref{eq:FODynamics}); these are the observed dynamics once the transients associated with the initial condition have died down. In particular, we consider systems whose post-transient dynamics are described by an attractor that is either a steady state, a periodic limit cycle or a quasi-periodic limit cycle, which encompasses the behavior of a large class of physical systems. The nature of the attractor is independent of the initial condition but depends on the value of $p$.

The purpose of the present work is to combine a ROM constructed using DMD and a Kalman filter to formulate an estimation algorithm that solves the following problem: given a sequence of measurements $\{\vy_1, \cdots, \vy_{k}\}$ from a post-transient reference trajectory of (\ref{eq:FODynamics}), estimate the high-dimensional state $\vz_{k}$ at current time $k$ without knowledge of $p$ itself. We will consider an adaptive approach whose goal is to estimate both the state and the parameter using an local dynamics model that depend on the current parameter estimate, and compare it to a robust approach whose goal is to estimate only the state $\vz_{k}$ using a global dynamics model, i.e., valid over an interval of $p$ values. In the remainder of this section, we describe the ROM methodologies forming the basis of these two approaches.
\section{REDUCED-ORDER MODELING}
\label{sec:ROM}
In order to make online estimation practical despite the high-dimensionality of (\ref{eq:FODynamics}), the first step is to formulate a reduced-order model (ROM) of the dynamics \cite{rowley2017}. Inspired by previous work \cite{taira2017,gomez2019data,tsolovikos2020estimation}, we construct a ROM by using the proper orthogonal decomposition (POD) to project the high-dimensional state in an appropriate reduced-order subspace, followed by applying the dynamic mode decomposition (DMD) to obtain a linear model for the dynamics within the subspace. Our novelty lie in the way that we account for multiple dynamical regimes as the physical parameter $p$ varies, resulting in two new methodologies to construct a ROM in the presence of parameter uncertainty.

Since both the POD and DMD are data-driven, we begin with a training data set of trajectories obtained by solving (\ref{eq:FODynamics}) for a range of values of $p$ specified by a finite set $\mathcal{P}_\mathrm{train} = \{p^{(1)}, \dots, p^{(q)}\} \subset [p_\mathrm{min},p_\mathrm{max}]$, resulting in a concatenated collection of snapshots $\mZ_\mathrm{train} = \{\mZ^{(i)}\}_{i=1}^q$, where each $\mZ^{(i)} = \{\vz_0^{(i)}, \dots, \vz_m^{(i)}\}$ is a post-transient trajectory of (\ref{eq:FODynamics1}) for a specific value $p^{(i)} \in \mathcal{P}_\mathrm{train}$. We arrange the concatenated snapshots into two time-shifted matrices
\begin{subequations}
\begin{align}
\mX &= \{\vz_0^{(1)}, \dots, \vz_{m-1}^{(1)}, \dots, \vz_0^{(q)}, \dots, \vz_{m-1}^{(q)}\}, \\
\mY &= \{\vz_1^{(1)}, \dots, \vz_m^{(1)}, \dots, \vz_1^{(q)}, \dots, \vz_m^{(q)}\}.
\end{align}
\label{eq:SnapshotMatrices}
\end{subequations}

To reduce the dimensionality of the system, we follow the POD methodology and perform a reduced-rank SVD of the data matrix $\mX$; that is, $\mX \simeq \mU \boldsymbol{\Sigma} \mV^\mathsf{T}$, where $\mU, \mV \in \mathbb{R}^{n \times r}$ are orthogonal, $\boldsymbol{\Sigma} \in \mathbb{R}^{r \times r}$ is diagonal, and $r$ is the rank of the truncation. The columns of $\mU$, called the POD modes, contain the $r$ most energetic spatial structures in the data (in an $L_2$ sense) and therefore span an energy-optimal subspace in which to project $\vz_k$.  The projection is defined by
\begin{equation}
\vx_k = \mU^\mathsf{T} \vz_k,
\label{eq:ReducedState}
\end{equation}
yielding a reduced-order state $\vx_k$ representing the subspace coordinates, or modal amplitudes, of $\vz_k$. Conversely, $\vz_k$ can be approximately recovered from $\vx_k$ as
\begin{equation}
\vz_k \simeq \mU \vx_k,
\label{eq:FullState}
\end{equation}
where equality holds only when $\vz_k$ is in the range of $\mU$.

To find a model for the dynamics of $\vx_k$, which is vastly cheaper to evolve than (\ref{eq:FODynamics}) when $r \ll n$, we employ the DMD, a purely data-driven algorithm that has found numerous applications in various fields \cite{schmid2010,kutz2016}. In its standard formulation, the DMD seeks a best-fit linear model of the dynamics for a single parameter $p^{(i)}$ in the form of a matrix $\mA \in \mathbb{R}^{n \times n}$ such that $\vz_{k+1}^{(i)} \simeq \mA \vz_k^{(i)}$ for all $k$, using the snapshots in $\mZ^{(i)}$. Projecting this matrix to the columns of $\mU$ results in a linear ROM $\vx_{k+1} \simeq \mA_r \vx_k$, where $\mA_r = \mU^\mathsf{T} \mA \mU  \in \mathbb{R}^{r \times r}$. Since this ROM is only trained using data from a single $p^{(i)}$, it will produce inaccurate dynamics for other parameter values. Thus, we introduce two variations of DMD in order to take into account the dependence of the dynamics on the physical parameter $p$.

\subsection{Robust reduced-order model (rROM)}

In the robust approach, we seek a ROM that approximately describes the dynamics corresponding to any \textcolor{black}{parameter value} $p \in [p_\mathrm{min},p_\mathrm{max}]$ but without explicit dependence on $p$. To do so, we compute a single matrix $\mA \in \mathbb{R}^{n \times n}$ such that $\vz_{k+1}^{(i)} \simeq \mA \vz_k^{(i)}$ for all $k = 0, \dots, m-1$ and for all $p^{(i)} \in \mathcal{P}_\mathrm{train}$. The best-fit linear model is given by $\mA = \mY \mX^\dagger$, where $\mX$ and $\mY$ are the matrices in (\ref{eq:SnapshotMatrices}) and $\mX^\dagger$ is the pseudoinverse of $\mX$. 

The relation (\ref{eq:FullState}) and the orthogonality of $\mU$ then yield a linear discrete-time robust ROM of the form
\begin{subequations}
\begin{align}
\vx_{k+1} &= \mA_r\vx_{k} + \vw_{k}, \label{eq:rROMDynamics} \\
\vy_k &= \vh_r(\vx_k, \vv_k), \label{eq:rROMObservation}
\end{align} \label{eq:rROM}%
\end{subequations}
where $\mA_r = \mU^\mathsf{T} \mA \mU  \in \mathbb{R}^{r \times r}$ is the reduced state-transition model and $\vh_r: \mathbb{R}^r \times \mathbb{R}^m \rightarrow \mathbb{R}^m$ is the reduced map from reduced state and noise to measurement, with $\vh_r(\vx_k, \vv_k) = \vh(\mU \vx_k, \vv_k)$. The non-Gaussian process noise $\vw_k$ and observation noise $\vv_k$ account for the POD modes left out of the truncated SVD of $\mX$, as well as the error incurred by the linear approximation and effective averaging of the dynamics over the parameter range $[p_\mathrm{min},p_\mathrm{max}]$. Finally, $\vz_k$ is recovered from $\vx_{k}$ using (\ref{eq:FullState}). Note that  $\mA_r$ can be directly calculated as $\mA_r = \mU^\mathsf{T} \mY \mV \boldsymbol{\Sigma}^{-1}$, which avoids forming the large $n \times n$ matrix $\mA$.

\subsection{Adaptive polytopic reduced-order model (apROM)}

\begin{figure}
\centering
\includegraphics[width=\linewidth]{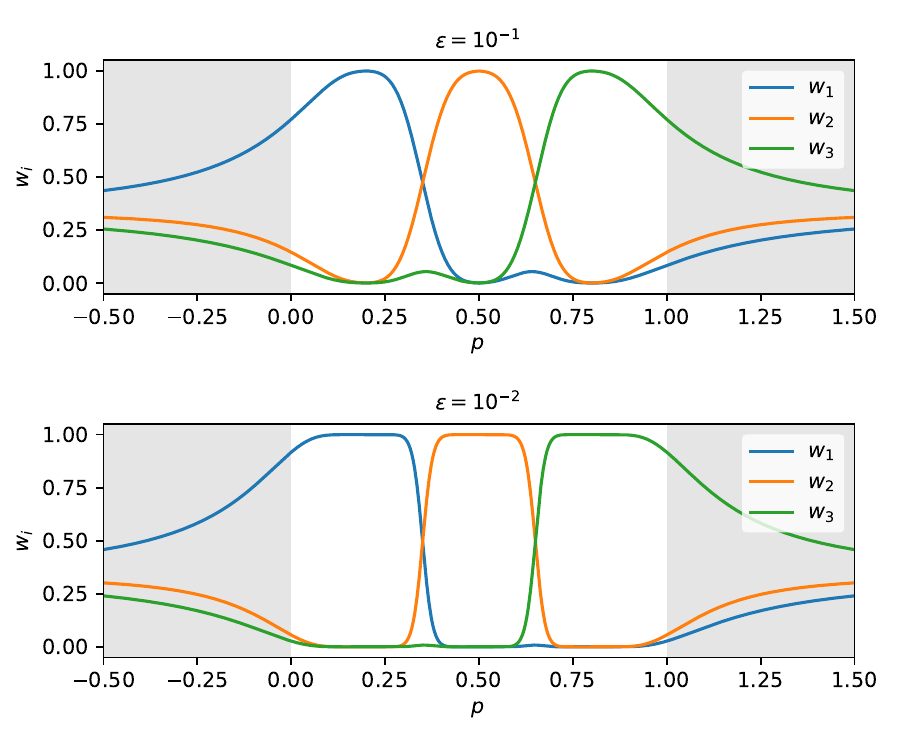}
\caption{Behavior of the weights $w_i(p)$ versus parameter $p$ for different values of $\epsilon$ in the case where $p_\mathrm{min} = 0$, $p_\mathrm{max} = 1$, $\mathcal{P}_\mathrm{train} = \{p^{(1)} = 0.2, p^{(2)} = 0.5, p^{(3)} = 0.8\}$.}
\label{fig:Weights}
\end{figure}

In the adaptive approach, we seek a ROM that better approximates the dynamics corresponding to various $p \in [p_1,p_2]$ by introducing an explicit dependence on $p$. For the purpose of formulating the ROM, $p$ is assumed known but it will later be estimated as part of the estimation algorithm. We first construct a library of matrices $\{\mA^{(1)}, \dots, \mA^{(q)}\}$ by separately applying DMD to each trajectory $\mZ^{(i)}$, for all $p^{(i)} \in \mathcal{P}_\mathrm{train}$. Thus, for each $p^{(i)}$ we compute $\mA^{(i)} \in \mathbb{R}^{n \times n}$ such that $\vz_{k+1}^{(i)} \simeq \mA^{(i)} \vz_k^{(i)}$ for all $k = 0, \dots, m-1$. The best-fit linear model is given by $\mA^{(i)} = \mY^{(i)} (\mX^{(i)})^\dagger$, where $\mX^{(i)}$ and $\mY^{(i)}$ are given by
\begin{subequations}
\begin{align}
\mX^{(i)} &= \{\vz_0^{(i)}, \dots, \vz_{m-1}^{(i)}\}, \\
\mY^{(i)} &= \{\vz_1^{(i)}, \dots, \vz_m^{(i)}\}.
\end{align}
\end{subequations}
We then construct an adaptive weighted average of these local linear models as
\begin{equation}
\bar{\mA}(p) = \sum_{i=1}^q w_i(p) \mA^{(i)},
\end{equation}
where the weights depend on $p$ through functions $w_i(p)$ that need to satisfy two properties: first, all $w_i(p)$'s must always sum to one; second, each $w_i(p)$ must increase monotonically with decreasing $|p-p^{(i)}|$. Thus, a possible choice is to define the weights using a `softmax' function as
\begin{subequations}
\begin{align}
w_i(p) &= \frac{e^{l_i(p)}}{\sum_{j=1}^q e^{l_j(p)}}, \\
l_i(p) &= \frac{1}{\epsilon+|p-p^{(i)}|/\Delta p},
\end{align}
\end{subequations}
where $\Delta p = p_\mathrm{max} - p_\mathrm{min}$, and $\epsilon > 0$ is a small number that prevents a division by zero and additionally serves as a control knob for the `sharpness' of each weight function $w_i(p)$ in the neighborhood of $p^{(i)}$; see Figure \ref{fig:Weights}. In the examples to follow, we choose $\epsilon = 10^{-2}$.

%%%%
%%%%

The relation (\ref{eq:FullState}) and the orthogonality of $\mU$ can then be applied to yield a linear discrete-time adaptive polytopic ROM of the form
\begin{subequations}
\begin{align}
\vx_{k+1} &= \bar{\mA}_r(p) \vx_{k} + \vw_{k}, \label{eq:aROMDynamics} \\
\vy_k &= \vh_r(\vx_k, \vv_k), \label{eq:aROMObservation}
\end{align} \label{eq:aROM}%
\end{subequations}
where $\bar{\mA}_r(p) = \mU^\mathsf{T} \bar{\mA}(p) \mU  \in \mathbb{R}^{r \times r}$ is an adaptive reduced state-transition model and $\vh_r: \mathbb{R}^r \times \mathbb{R}^p \rightarrow \mathbb{R}^p$ is the same map as in (\ref{eq:rROMObservation}). As in the previous case of the rROM, the non-Gaussian process noise $\vw_k$ and observation noise $\vv_k$ account for the POD modes left out of the truncated SVD of $\mX$, as well as the error incurred by the linear approximation. However, the dependence of the dynamics on $p$ is now expressed by the weights $w_i(p)$, which push $\bar{\mA}_r(p)$ towards the matrix $\mA^{(i)}$ corresponding to the parameter $p^{(i)}$ that is closest to $p$. As before, $\vz_k$ is recovered from $\vx_{k}$ using (\ref{eq:FullState}).

\section{ESTIMATION ALGORITHM} 
\label{sec:estimation}
We now propose two state estimation strategies in the presence of parameter uncertainty: first, an adaptive modular parameter-state estimation algorithm that combines the adaptive polytopic ROM with two Kalman filters to estimate separately the state and the unknown parameter; second, an adaptive joint parameter-state estimation algorithm that combines the adaptive polytopic ROM with a single Kalman filter to estimate jointly the state and the unknown parameter. Finally, we contrast the proposed adaptive estimators with a classical robust state estimation algorithm that combines the robust ROM with a single Kalman filter to provide an estimate for the state.

\subsection{Adaptive modular parameter-state estimation (apROM-mKF)}

To estimate both the reduced state $\vx_k$ and model parameter $p$ from the noisy measurement data, we first consider a modular dual parameter-state estimation algorithm inspired by the dual extended and unscented Kalman filters \cite{wan2000unscented,wan2001dual,wenzel2006dual}. In this approach, we simultaneously run two unscented Kalman filters in parallel, one to provide a state estimate $\hat{\vx}_k$ and one to provide a parameter estimate $\hat{p}_k$. The state-space representation for the state is given by the adaptive polytopic ROM (\ref{eq:aROM}), while that for the parameter is given by
\begin{subequations}
\begin{align}
p_{k+1} &= p_{k}, \\
\vy_k &= \vh_r(\bar{\mA}_r(p_k) \vx_{k-1}, \vv_k). \label{eq:parameterROMObservation}
\end{align} \label{eq:parameterROM}%
\end{subequations}
When running simultaneously the state and parameter filters, we use at each time step the state estimate in the parameter filter, and the parameter estimate in the state filter.

\subsection{Adaptive joint parameter-state estimation (apROM-jKF)}

The second approach we propose to estimate both the reduced state $\vx_k$ and model parameter $p$ is a joint parameter-state estimation algorithm inspired by the joint extended and unscented Kalman filters \cite{wan2000unscented}. In this approach, we construct a joint state vector $\vx'_k = [\vx_k^\mathsf{T}, p_k^\mathsf{T}]^\mathsf{T}$. The state-space representation for this joint state is given by a combination of the adaptive ROM (\ref{eq:aROM}) and the parameter equation (\ref{eq:parameterROM}), that is,
\begin{subequations}
\begin{align}
\begin{bmatrix} \vx_{k+1} \\ p_{k+1} \end{bmatrix} &= \begin{bmatrix} \bar{\mA}_r(p) \vx_{k} \\ p_k \end{bmatrix} + \begin{bmatrix} \vw_k \\ 0 \end{bmatrix}, \\
\vy_k &= \vh_r(\vx_k, \vv_k).
\end{align}%
\end{subequations}
We can then run a single unscented Kalman filter on the joint state vector to recover an estimate for both the state and the parameter.

\subsection{Robust parameter estimation (rROM-KF)}
For our baseline, we consider a robust state estimation algorithm formed by combining the robust ROM (\ref{eq:rROM}) with an unscented Kalman filter, or a simple vanilla Kalman filter if the observation model is linear.

\section{RESULTS}
\label{sec:Simulations}
We evaluate the state estimation performance of the proposed adaptive parameter-state estimation algorithms and compare them with the robust state estimation method, for systems governed by the nonlinear Burgers equation and the Navier-Stokes equations in the presence of parameter uncertainty. In all cases, we initialize the state estimate to zero, and the parameter estimate to the mean of the values in the training set $\mathcal{P}_\mathrm{train}$.

\subsection{Burgers equation}
\label{sec:Burgers}

The forced Burgers equation is a prototypical nonlinear hyperbolic PDE that takes the form
\begin{equation}
\frac{\partial u}{\partial t} + u \frac{\partial u}{\partial x} - \nu \frac{\partial^2 u}{\partial x^2} = f(x,t), 
\label{eq:Burgers}
\end{equation}
where $u(x,t)$ is the velocity at position $x \in [0,L]$ and time $t$, $f(x,t)$ is a distributed time-dependent forcing, and the scalar $\nu$ acts like a viscosity. Here, we choose a forcing of the form
\begin{equation}
f(x,t) = 2 \sin(\omega t - k x),
\end{equation}
where $k = 2 \pi/L$, and we let $\nu$ and $\omega$ be related through a scalar parameter $p \in [0,1]$ as follows:
\begin{subequations}
\begin{align}
\nu &= \nu_1 + (\nu_2-\nu_1) p, \\
\omega &= \omega_1 + (\omega_2-\omega_1) p.
\end{align}%
\end{subequations}
Thus, $p$ can be regarded as a physical parameter that affects the dynamics of the forced Burgers equation through both $\nu$ and $\omega$.  We consider periodic boundary conditions and choose $L=1$, $\nu_1 = 0.01$, $\nu_2 = 0.1$, $\omega_1 = 0.2 \pi$, $\omega_2 = 0.4 \pi$.

We solve the forced Burgers equation using a spectral method with $n = 256$ Fourier modes and a fifth-order Runge-Kutta time integration scheme. We define the discrete-time state vector $\vz_k \in \mathbb{R}^n$ that contains the values of $u$ at $n$ equally-spaced collocation points and at discrete time steps $t = k \Delta t$, where $\Delta t = 0.05$. To generate the training dataset $\mZ_\mathrm{train}$ used for constructing the ROM, we compute solutions of the Burgers equation corresponding to $p^{(i)} \in \mathcal{P}_\mathrm{train} = \{0,0.2,0.4,0.6,0.8,1\}$. For each $p^{(i)}$, we discard the transient portion of the dynamics and save 2001 snapshots $\mZ^{(i)} = \{\vz_0^{(i)}, \dots, \vz_{2000}^{(i)}\}$ in the post-transient regime. We retain $r = 10$ modes when constructing the ROM, corresponding to an-order-of-magnitude reduction in the dimensionality of the system. In the following examples, we consider measurement data obtained from 4 sensors equally-spaced along the physical domain and polluted with white noise of standard deviation 0.05.

\begin{figure}
\centering
\includegraphics[width=\linewidth]{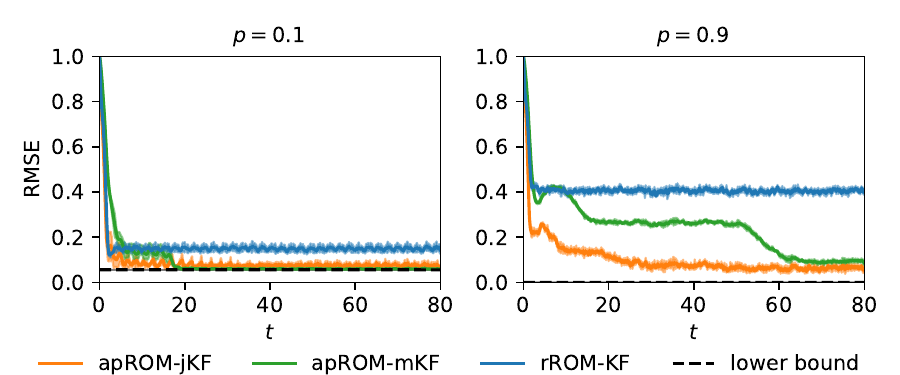}
\caption{Relative mean square error (RMSE) of the state estimate for different values not included in $
\mathcal{P}_\mathrm{train}$ of the unknown parameter $p$, for Burgers' equation. The simulations are repeated for 5 initial conditions of the reference state; the shaded area indicates the standard deviation of the RMSE. The lower bound denotes the error incurred by the projection to the POD modes.}
\label{fig:BurgersError}
\end{figure}

\begin{figure}
\centering
\includegraphics[width=\linewidth]{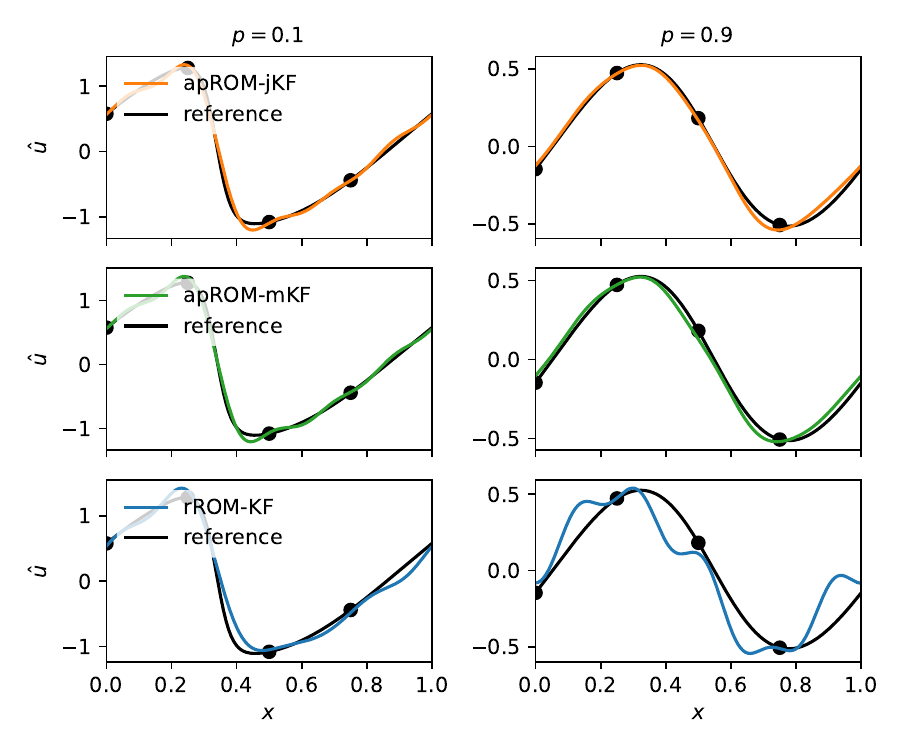}
\caption{State estimate $\hat{u}(x,t)$ at final time for different values not included in $
\mathcal{P}_\mathrm{train}$ of the unknown parameter $p$, for Burgers' equation.}
\label{fig:BurgersStateEstimate}
\end{figure}

\begin{figure}
\centering
\includegraphics[width=\linewidth]{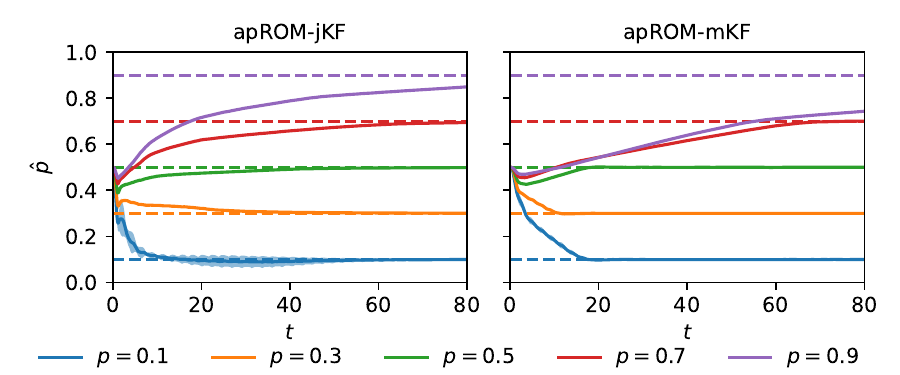}
\caption{Parameter estimate $\hat{p}$ for different values not included in $
\mathcal{P}_\mathrm{train}$ of the unknown parameter $p$, for Burgers' equation. The (unknown) true parameter values are shown in dotted lines. The simulations are repeated for 5 initial conditions of the reference state; the shaded area indicates the standard deviation of the RMSE.}
\label{fig:BurgersParamEstimate}
\end{figure}

\begin{figure}
\centering
\includegraphics[width=\linewidth]{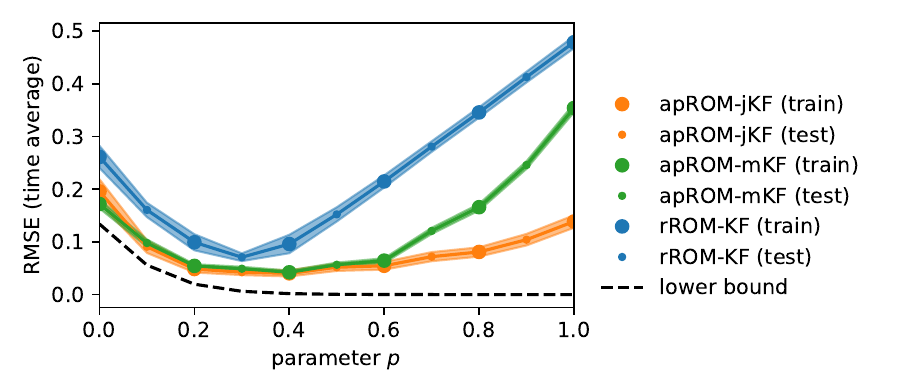}
\caption{Relative mean square error (RMSE) of the state estimate averaged over time for parameter values included in the training data and not included in the training data (denoted test) of the unknown parameter $p$, for Burgers' equation. The simulations are repeated for 5 initial conditions of the reference state; the shaded area indicates the standard deviation of the RMSE. The lower bound denotes the error incurred by the projection to the POD modes.}
\label{fig:BurgersErrorFinalTime}
\end{figure}

We now utilize the adaptive and robust filters to estimate the states using measurements coming from systems with parameter values not contained in the training data. Figures \ref{fig:BurgersError} and \ref{fig:BurgersStateEstimate} demonstrate the superior estimation accuracy with the adaptive parameter-state estimation approaches (apROM-jKF and apROM-mKF) compared to the robust state estimation approach (rROM-KF). Furthermore, Figure \ref{fig:BurgersParamEstimate} shows that the adaptive approaches yield an accurate estimate of the unknown model parameter for several parameter values. Finally, Figure \ref{fig:BurgersErrorFinalTime} shows the time-averaged estimation error for parameter values both included and not included in the training data. Generally, parameter values closer to the ends of the training set $\mathcal{P}_\mathrm{train}$ lead to worse errors since they take longer to converge, as evidenced in Figure \ref{fig:BurgersParamEstimate}.

%%%%
%%%%

%%%%
%%%%

%%%%
%%%%

%%%%
%%%%

\subsection{Navier-Stokes equations}
\label{sec:Navier-Stokes}

The Navier-Stokes equations are a set of nonlinear PDEs that describe the motion of fluids flows. For incompressible fluids, the Navier-Stokes equations take the form
\begin{subequations}
\begin{align}
\frac{\partial \vu}{\partial t} + (\vu \cdot \nabla) \vu &= - \nabla P + \frac{1}{Re} \Delta \vu, \\
\nabla \cdot \vu = 0,
\end{align} \label{eq:NS}%
\end{subequations}
where $\vu(\vx,t)$ and $P(\vx,t)$ are the velocity vector and pressure at position $\vx$ and time $t$, and the scalar $Re$ is the Reynolds number. We consider the classical problem of a flow past a cylinder in a 2D domain, which is well known to exhibit vortex shedding above a critical Reynolds number $Re_c \sim 40$ \cite{jackson1987finite}. For our study, we focus on the range $Re \in [50,100]$. The shedding frequency and the spacing between consecutive vortices are both functions of $Re$.

\begin{figure}
\centering
\includegraphics[width=\linewidth]{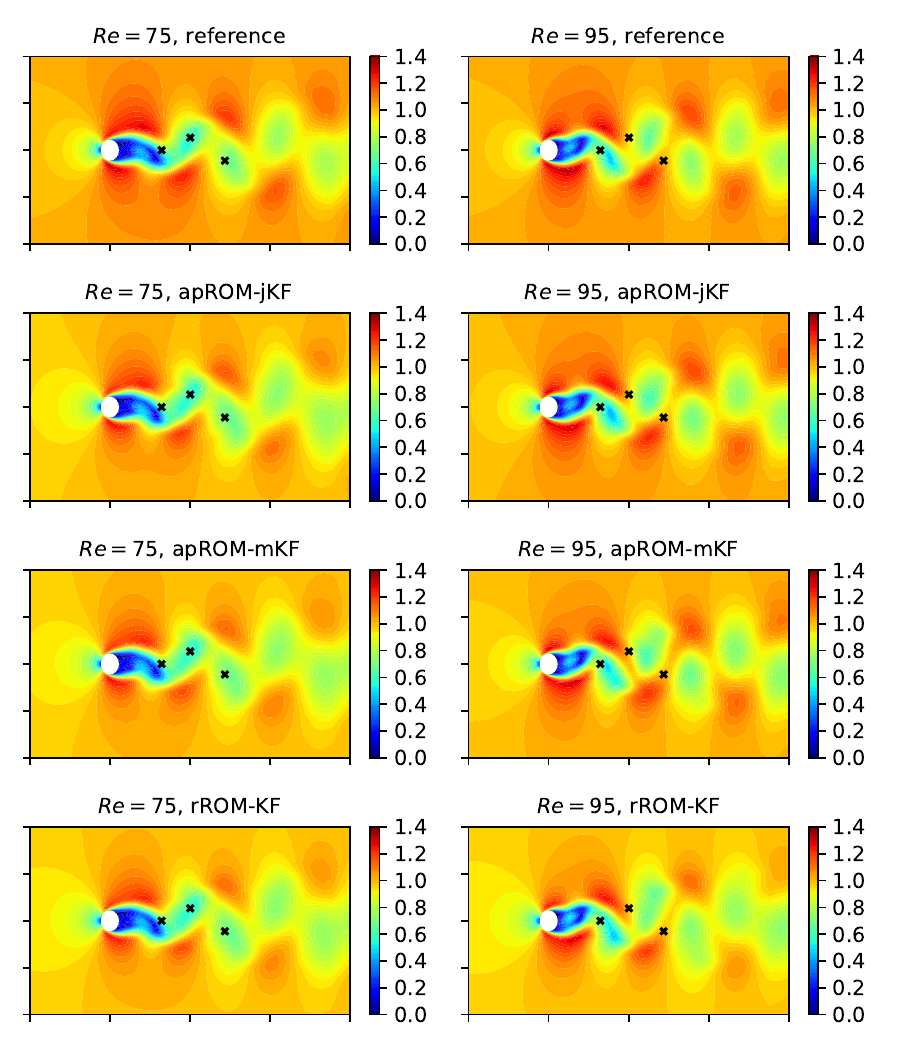}
\caption{State estimate $\hat{\vu}(\vx,t)$ at final time for different values not included in $\mathcal{P}_\mathrm{train}$ of the unknown parameter $Re$, for the Navier-Stokes equations.}
\label{fig:NSStateEstimate}
\end{figure}

We solve the Navier-Stokes equations with the open source finite volume code OpenFOAM using a mesh consisting of 18840 nodes and a second-order implicit scheme with time step 0.05. The discrete-time state vector $\vz_k \in \mathbb{R}^{37680}$ contains the two velocity components of $\vu$ at discrete time steps $t = k \Delta t$, where we choose $\Delta t = 0.25$. To generate the training dataset $\mZ_\mathrm{train}$ for constructing the ROM, we run simulations of the Navier-Stokes equations for $Re \in \mathcal{P}_\mathrm{train} = \{50,60,70,80,90,100\}$. We discard the transient portion of the dynamics and save 401 snapshots $\mZ^{(i)} = \{\vz_0^{(i)}, \dots, \vz_{400}^{(i)}\}$ in the post-transient regime. We retain $r = 20$ modes when constructing the ROM, corresponding to a three-orders-of-magnitude reduction in the dimensionality of the system. In the following examples, we consider measurement data obtained from $3$ sensors polluted with white noise of standard deviation 0.05. Each sensor measures both components of velocity, and their locations are indicated by the black crosses in Figure \ref{fig:NSStateEstimate}.

\begin{figure}
\centering
\includegraphics[width=\linewidth]{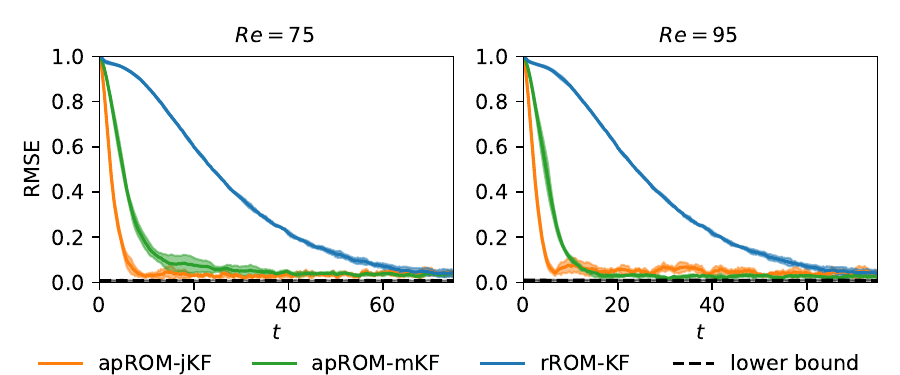}
\caption{Relative mean square error (RMSE) of the state estimate for different values not included in $
\mathcal{P}_\mathrm{train}$ of the unknown parameter $Re$, for the Navier-Stokes equations. The simulations are repeated for 5 initial conditions of the reference state; the shaded area indicates the standard deviation of the RMSE. The lower bound denotes the error incurred by the projection to the POD modes.}
\label{fig:NSError}
\end{figure}

\begin{figure}
\centering
\includegraphics[width=\linewidth]{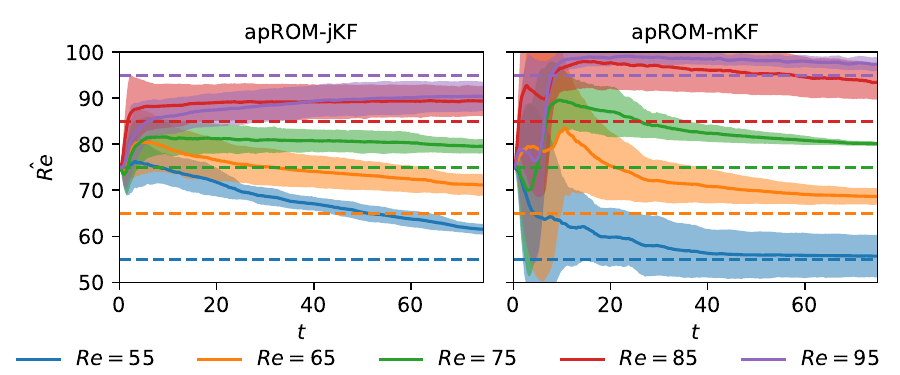}
\caption{Parameter estimate $\hat{Re}$ for different values not included in $
\mathcal{P}_\mathrm{train}$ of the unknown parameter $Re$, for the Navier-Stokes equations. The (unknown) true parameter values are shown in dotted lines. The simulations are repeated for 5 initial conditions of the reference state; the shaded area indicates the standard deviation of the RMSE.}
\label{fig:NSParamEstimate}
\end{figure}

\begin{figure}
\centering
\includegraphics[width=\linewidth]{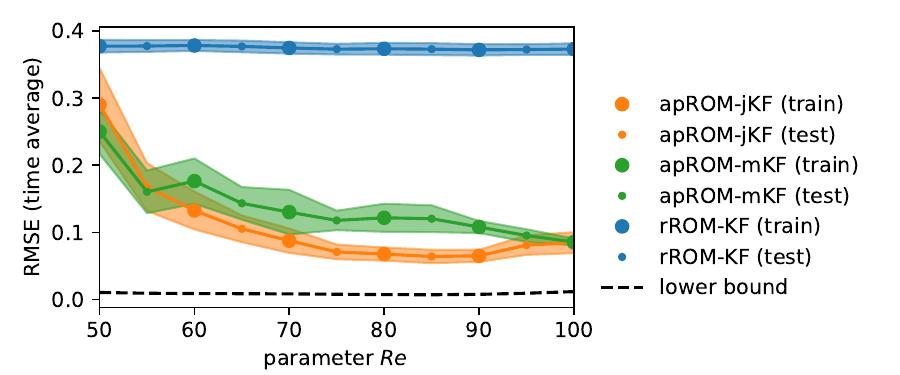}
\caption{Relative mean square error (RMSE) of the state estimate averaged over time for parameter values included in the training data and not included in the training data (denoted test) of the unknown parameter $Re$, for the Navier-Stokes equations. The simulations are repeated for 5 initial conditions of the reference state; the shaded area indicates the standard deviation of the RMSE. The lower bound denotes the error incurred by the projection to the POD modes.}
\label{fig:NSErrorFinalTime}
\end{figure}

We now utilize our filters to estimate the states using measurements coming from systems with parameter values not contained in the training data. Figures \ref{fig:NSError}, \ref{fig:NSStateEstimate}, and \ref{fig:NSErrorFinalTime} demonstrate the superior estimation accuracy with the adaptive parameter-state estimation approaches (apROM-jKF and apROM-mKF) compared to the robust state estimation approach (rROM-KF). Furthermore, Figure \ref{fig:NSParamEstimate} shows that the adaptive approaches yield an accurate estimate for the unknown model parameter for several values of the uncertainty.

%%%%
%%%%

%%%%
%%%%

%%%%
%%%%

%%%%
%%%%

\section{CONCLUSIONS}
We considered the problem of dual parameter-state estimation for systems modeled by PDEs with parameter uncertainty. We proposed two adaptive dual estimators based on a polytopic representation of the system constructed using a library of local DMD models. We compared these methods with a classical robust state estimator on the Burgers and Navier-Stokes equations. The results reported here are promising and warrant further investigations of these methods. For instance, while we have only considered cases where the true physical parameter does not change with time, we plan to investigate scenarios with slowly varying parameter value. We also plan to conduct further numerical tests on the Navier-Stokes equations in more challenging regimes involving higher Reynolds numbers and more complex dynamics.
\label{sec:Conclusions}

\addtolength{\textheight}{-12cm}   % This command serves to balance the column lengths
                                  % on the last page of the document manually. It shortens
                                  % the textheight of the last page by a suitable amount.
                                  % This command does not take effect until the next page
                                  % so it should come on the page before the last. Make
                                  % sure that you do not shorten the textheight too much.

%%%%%%%%%%%%%%%%%%%%%%%%%%%%%%%%%%%%%%%%%%%%%%%%%%%%%%%%%%%%%%%%%%%%%%%%%%%%%%%%

%%%%%%%%%%%%%%%%%%%%%%%%%%%%%%%%%%%%%%%%%%%%%%%%%%%%%%%%%%%%%%%%%%%%%%%%%%%%%%%%

%%%%%%%%%%%%%%%%%%%%%%%%%%%%%%%%%%%%%%%%%%%%%%%%%%%%%%%%%%%%%%%%%%%%%%%%%%%%%%%%

%%%%%%%%%%%%%%%%%%%%%%%%%%%%%%%%%%%%%%%%%%%%%%%%%%%%%%%%%%%%%%%%%%%%%%%%%%%%%%%%

\bibliography{IEEEabrv,bibliography}
\bibliographystyle{IEEEtran}

\end{document}